\documentclass[a4paper]{jpconf}
\usepackage{amsmath}
\usepackage{graphicx}

\begin{document}

\title{Recovering a spinning inspiralling compact binary waveform immersed in LIGO-like noise with spinning templates}
\author{L\'{a}szl\'{o} Ver\'{e}b$^{1,2}$, Zolt\'{a}n Keresztes$^{1,2\star }$%
, P\'{e}ter Raffai$^{3\dag }$, Szabolcs M\'{e}sz\'{a}ros$^{4}$, L\'{a}szl%
\'{o} \'{A}. Gergely$^{1,2,5\ddag }$}

\address{$^{1}$ Department of Theoretical Physics, University of Szeged, Tisza Lajos
krt 84-86, Szeged 6720, Hungary\\
$^{2}$ Department of Experimental Physics, University of Szeged, D\'{o}m t\'{e}r 9, Szeged 6720, Hungary\\
$^{3}$ Institute of Physics, Lor\'and E\"otv\"os University, P\'azmany P\'eter s 1/A, Budapest 1117, Hungary\\
$^{4}$ Department of Optics and Quantum Electronics, University of Szeged, D\'{o}m t\'{e}r 9, Szeged 6720, Hungary\\
$^{5}$ Institute for Advanced Study, Collegium Budapest, Szenth\'aroms\'ag u 2, Budapest 1014, Hungary}

\ead{$^{\ast}$zkeresztes@titan.physx.u-szeged.hu\quad $^{\dag}$praffai@bolyai.elte.hu
\quad $^{\ddag }$gergely@physx.u-szeged.hu}

\begin{abstract}
We investigate the recovery chances of highly spinning waveforms immersed
in LIGO S5-like noise by performing a matched filtering with $10^{6}$
randomly chosen spinning waveforms generated with the LAL package. While the
masses of the compact binary are reasonably well recovered (slightly
overestimated), the same does not hold true for the spins. We show the best
fit matches both in the time-domain and the frequency-domain. These
encompass some of the spinning characteristics of the signal, but far less
than what would be required to identify the astrophysical parameters of the
system. An improvement of the matching method is necessary, though may be
difficult due to the noisy signal.
\end{abstract}

We report on an attempt to recover a gravitational wave signal $h_{injection}
$ immersed in LIGO S5-like noise by a matched filtering with $10^{6}$
randomly chosen templates generated with spinning waveforms \cite{LAL}. 
We
applied the following process to generate a LIGO S5-like noise sample of N
elements: (i) cut out a stretch of 2N elements from an arbitrary sample of
real LIGO S5 noise, (ii) cut the stretch into half, resulting in two
stretches of N elements, (iii) Fourier-transfom both stretches with an FFT
method, resulting in two samples of N complex elements, (iv) combine the
magnitude values of the first spectrum with the phase values of the second
spectrum, resulting in a complex spectrum of N elements, (v) apply an
inverse FFT on the combined spectrum, and use the real part of the resulting
array as our LIGO S5-like noise sample. This method has the advantage of
preserving the basic spectral and statistical properties of the original
LIGO data, while discarding any real gravitational wave signatures~that
might be present within the original data stretch.
\begin{figure}[th]
\includegraphics[height=8cm, angle=270]{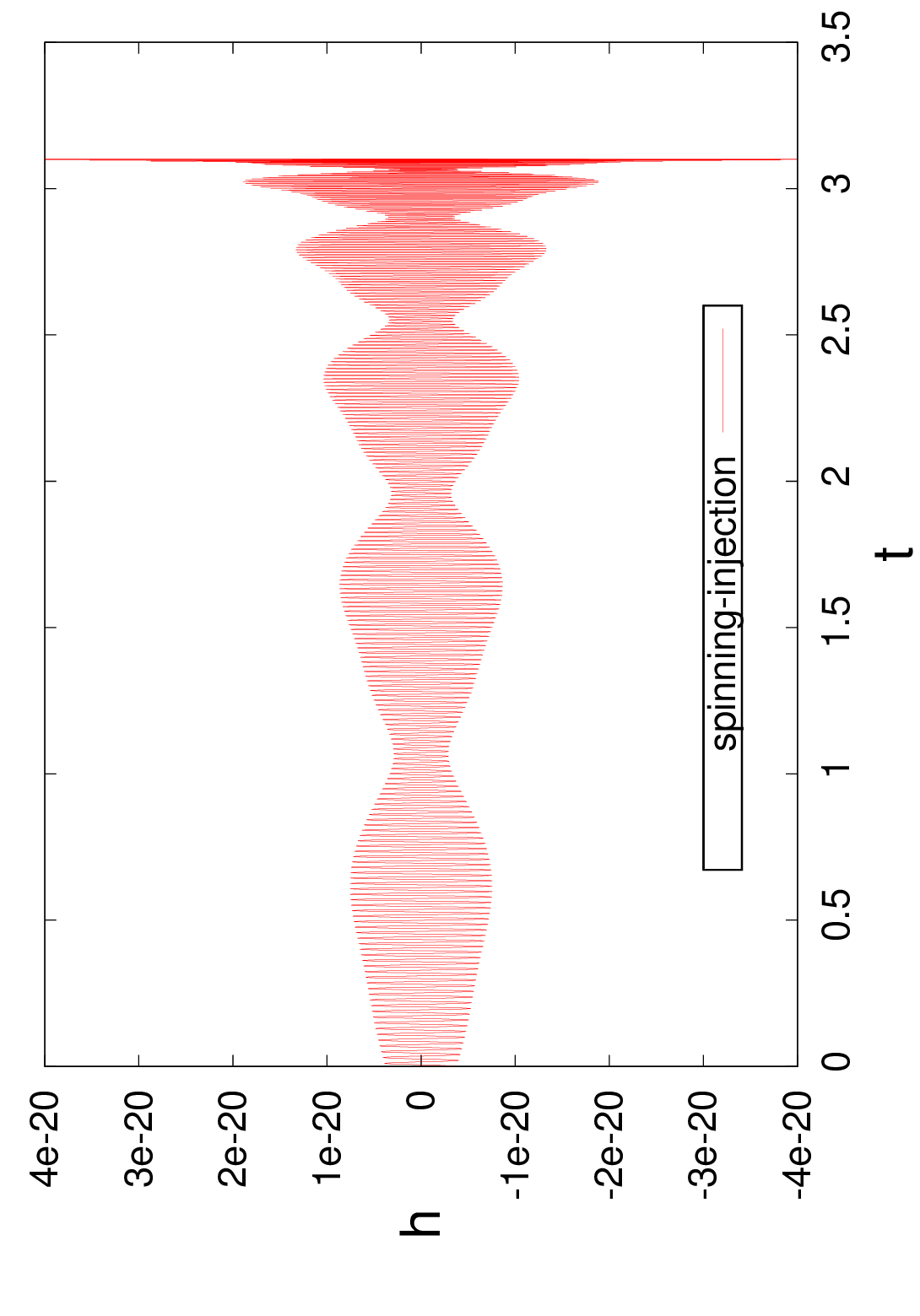} %
\includegraphics[height=8cm, angle=270]{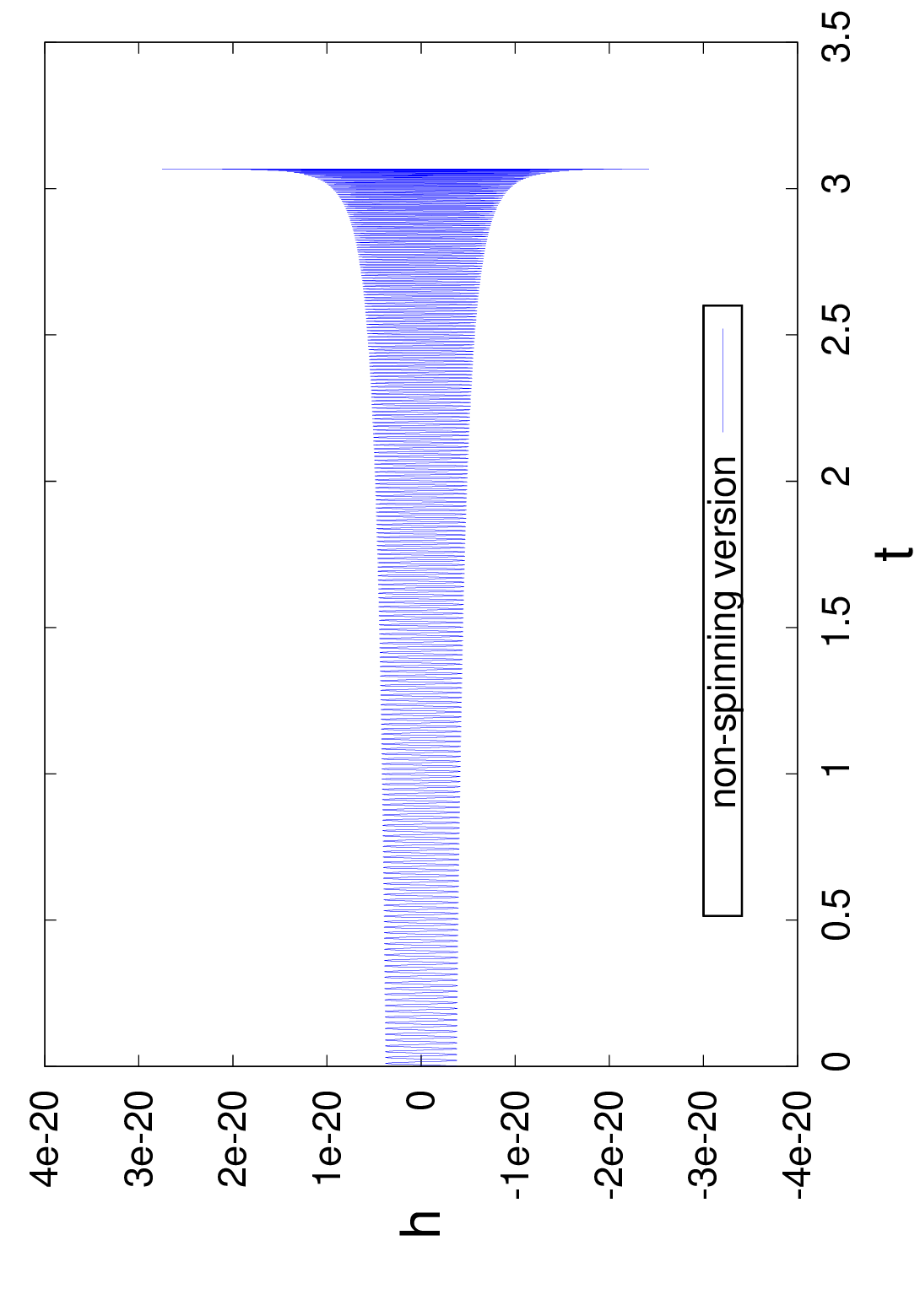}
\caption{ The injected signal (red) and its spinn-less counterpart (blue).}
\label{fig1}
\end{figure}

The matched filtering follows the method described in Ref. \cite{matching}.
Namely, we calculate the \textit{overlap} between the noisy injection $%
h_{n,i}$ and the spinning templates $h_{template}$ as 
\begin{equation}
O\left[ h_{n,i},h_{template}\right] =\frac{\left\langle
h_{n,i}|h_{template}\right\rangle }{\sqrt{\left\langle
h_{n,i}|h_{n,i}\right\rangle \left\langle
h_{template}|h_{template}\right\rangle }}~,  \label{O}
\end{equation}%
where%
\begin{equation}
\left\langle h_{1}|h_{2}\right\rangle =4\mathit{Re}\int_{f_{\min }}^{f_{\max
}}\frac{\tilde{h}_{1}\left( f\right) \tilde{h}_{2}^{\ast }\left( f\right) }{%
S_{n}\left( f\right) }df~,
\end{equation}%
with $\tilde{h}_{i}\left( f\right) $ the Fourier transform of $h_{i}\left(
t\right) $ and $S_{n}\left( f\right) $ the power spectral density of the
noise. We choose $f_{\min }=50$Hz and $f_{\max }=600$Hz, which lies in the
best sensitivity of the LIGO detectors, also the post-Newtonian prediction
for the waveform can be considered reasonably accurate. The most lengthy
templates in this domain were found to last for approximately $3$ sec. 
\begin{figure}[th]
\centerline{\includegraphics[height=5.5cm, angle=0]{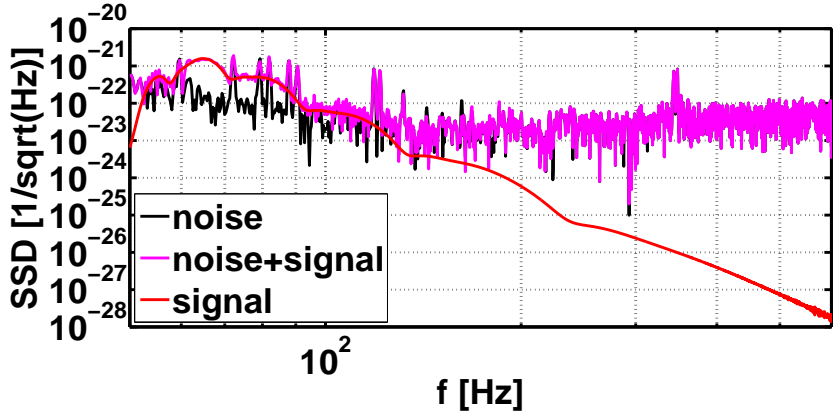}}
\caption{ The strain spectral density of the signal (red), noise (grey) and
injected signal + noise (magenta).}
\label{fig2}
\end{figure}

On theoretical grounds, an infinitely long data series would be required for
exact determination of the power spectrum. In order to achieve a stability
of the power spectrum of the order of $1\%$, we would like to have at least $%
100$ periods of all frequencies to be included in the power spectrum,
therefore a minimal length of the templates of $100\times \left( 1/50\text{Hz%
}\right) =2$ sec was imposed. As we know where the injection commences in
the noise time sequence, we simplify our analysis by comparing with
templates beginning at the same instant $t_{0}$. Further, we also assume
identical antenna functions of $\sqrt{2}/2$ both for the injection and
templates. Moreover, all parameters of the injection and templates other
then the masses $m_{i}$ and spins (given by their dimensionless magnitude $%
\chi _{i}$, polar and azimuthal angles $\beta _{i}$, $\varphi _{i}$, taken
in the inertial system with the line of sight on the $z$-axis) were kept
fixed. These parameters were the distance of the binary $d=1$Mpc, and the
polar angle of the orbital angular momentum $\Theta =1.4299$. The initial
phase $\Phi _{S}=0$. 

We picked up the injection signal as one of the longests available (see Fig %
\ref{fig1}). The presence of a high spin induces a strong amplitude
modulation, but also a less visible frequency modulation. For comparison, we
have also represented on Fig \ref{fig1} the corresponding non-spinning
signal, with the same masses, but the spins switched off. 
\begin{figure}[th]
\includegraphics[height=15cm, angle=270]{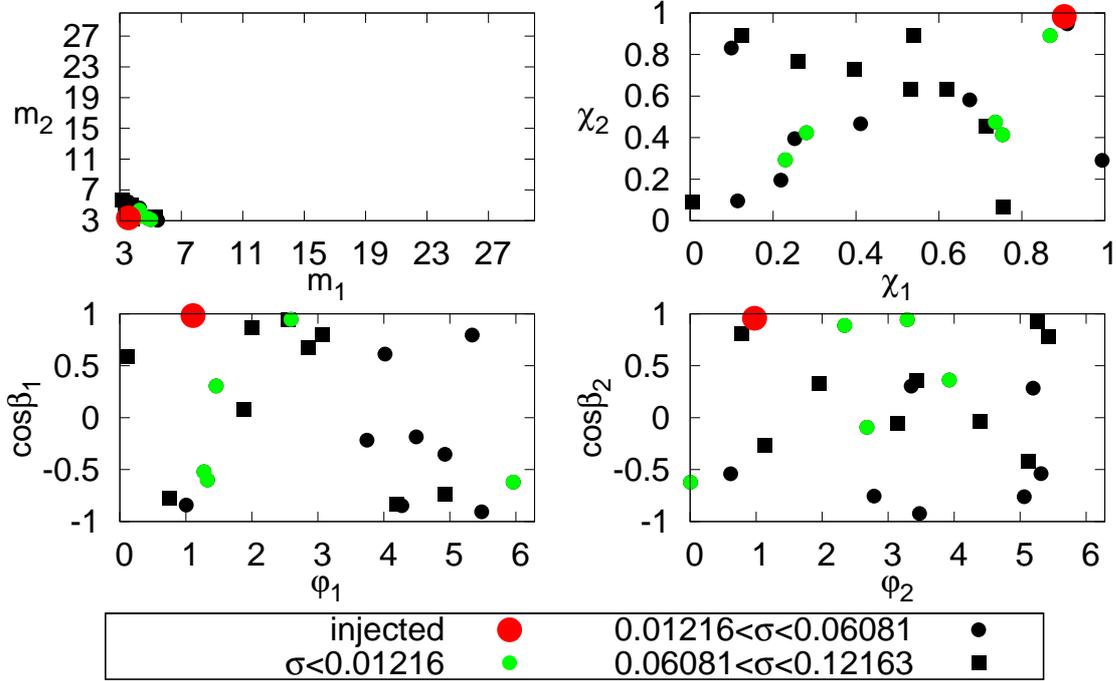}
\caption{ The injection and the best fit matches from $10^{6}$ templates.
The mass is roughly recovered, though sligtly overestimated, however the
information contained in the spins is not.}
\label{fig3}
\end{figure}
\begin{table}[t]
\caption{The parameters of the injection and 5 best fit templates.}
\label{table1}$%
\begin{tabular}{c|c|c|c|c|c|c|c|c}
name & $m_{1}(M_{\odot })$ & $m_{2}(M_{\odot })$ & $\chi _{1}$ & $\chi _{2}$
& $\cos \beta _{1}$ & $\varphi _{1}$ & $\cos \beta _{2}$ & $\varphi _{2}$ \\ 
\hline
injection & $3.553$ & $3.358$ & $0.983$ & $0.902$ & $0.984$ & $1.109$ & $%
0.978$ & $0.957$ \\ 
560884.wave & $4.805$ & $3.336$ & $0.230$ & $0.293$ & $-0.599$ & $1.327$ & $%
-0.621$ & $0.006$ \\ 
467403.wave & $4.092$ & $3.405$ & $0.868$ & $0.891$ & $-0.520$ & $1.275$ & $%
0.363$ & $3.925$ \\ 
313940.wave & $3.569$ & $3.540$ & $0.753$ & $0.413$ & $0.306$ & $1.458$ & $%
-0.092$ & $2.679$ \\ 
93938.wave & $5.012$ & $3.152$ & $0.280$ & $0.424$ & $0.947$ & $2.593$ & $%
0.888$ & $2.340$ \\ 
80521.wave & $4.278$ & $4.360$ & $0.736$ & $0.475$ & $-0.620$ & $5.959$ & $%
0.945$ & $3.290$%
\end{tabular}%
$%
\end{table}

The spinning signal is injected in the middle of a finite stretch of
LIGO-like data with double the length of the signal. The noise has the basic
statistical characteristics (e.g. amplitude distribution, spectral
compatibility) of the LIGO 4 km detectors during the 5th science run. The
segment to be analyzed is filtered with a zero-phase $50-600\ \mathrm{Hz}$
bandpass filter, and a zero-phase notch filter mitigating the effect of
narrow insensitivity peaks of detector spectrum. After removing one-fourth
of the data segment from both ends due to filter transients, this leaves
half the length of the initial noise sequence including the signal for
further processing. The strain spectral density, defined as the square root
of the periodogram, computed under MatLab \cite{SPS}, is represented on Fig %
\ref{fig2} for the signal, noise and noisy signal. 
\begin{figure}[th]
\includegraphics[height=8cm, angle=270]{injected.ps} %
\includegraphics[height=8cm, angle=270]{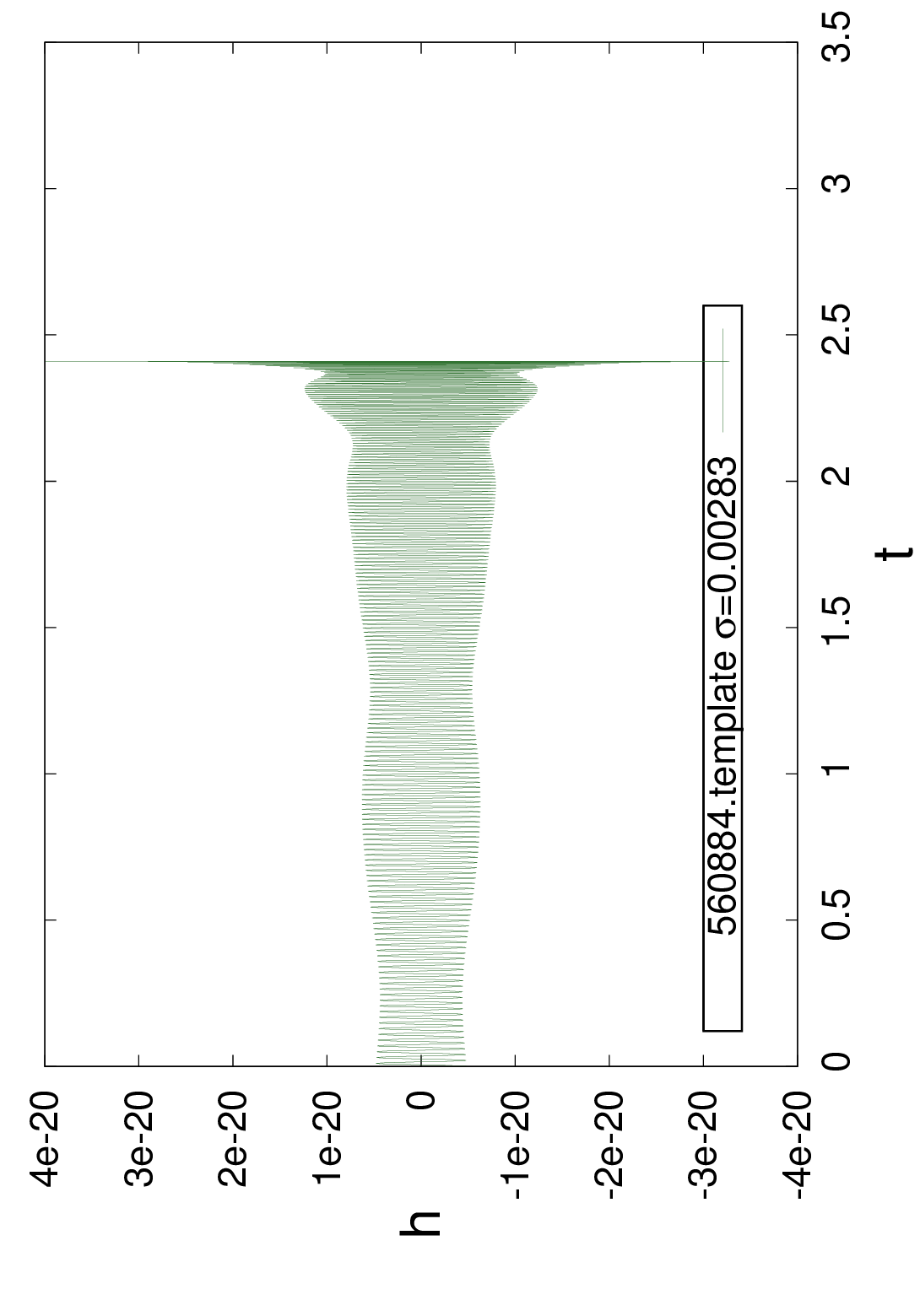} %
\includegraphics[height=8cm, angle=270]{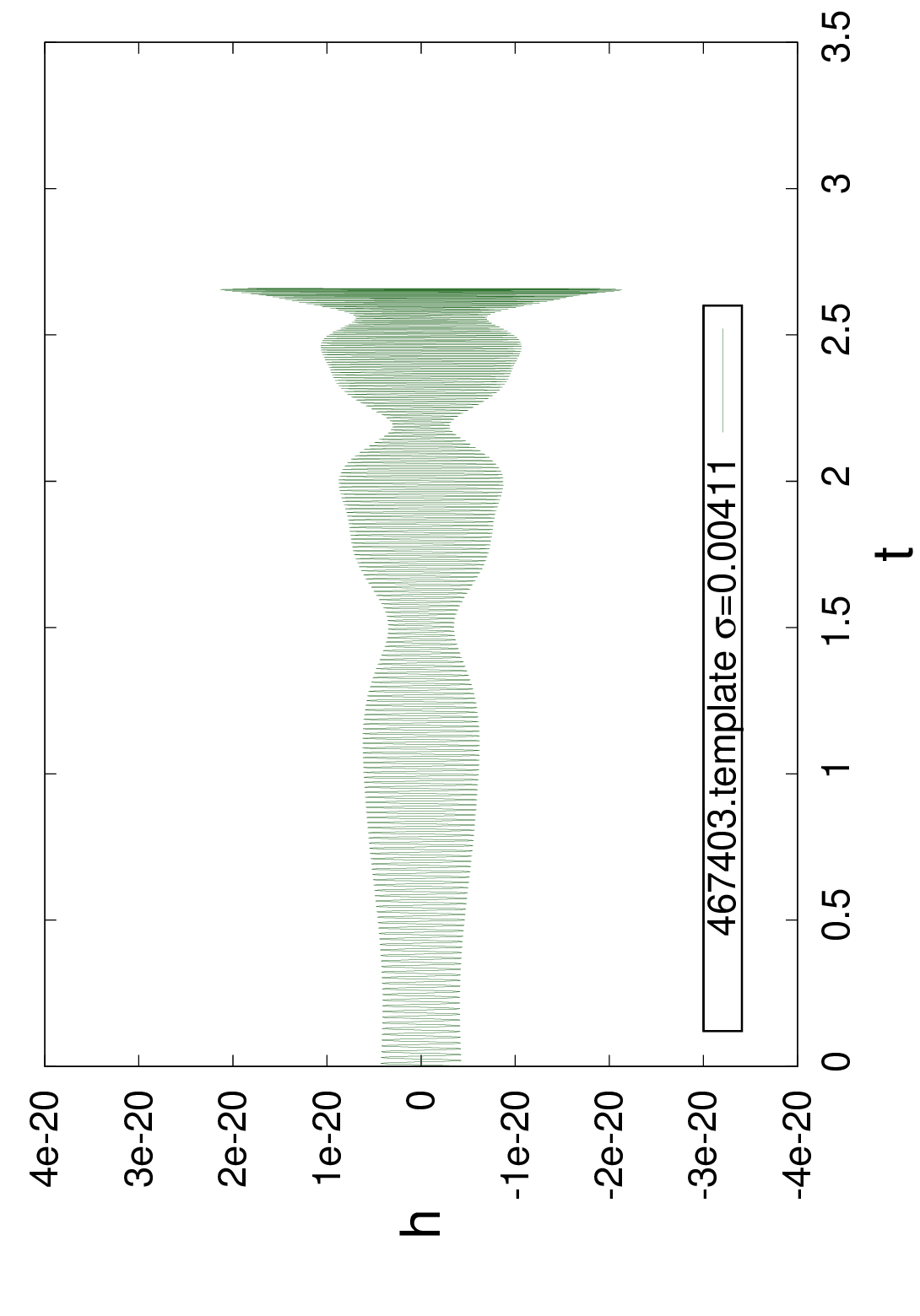} %
\includegraphics[height=8cm, angle=270]{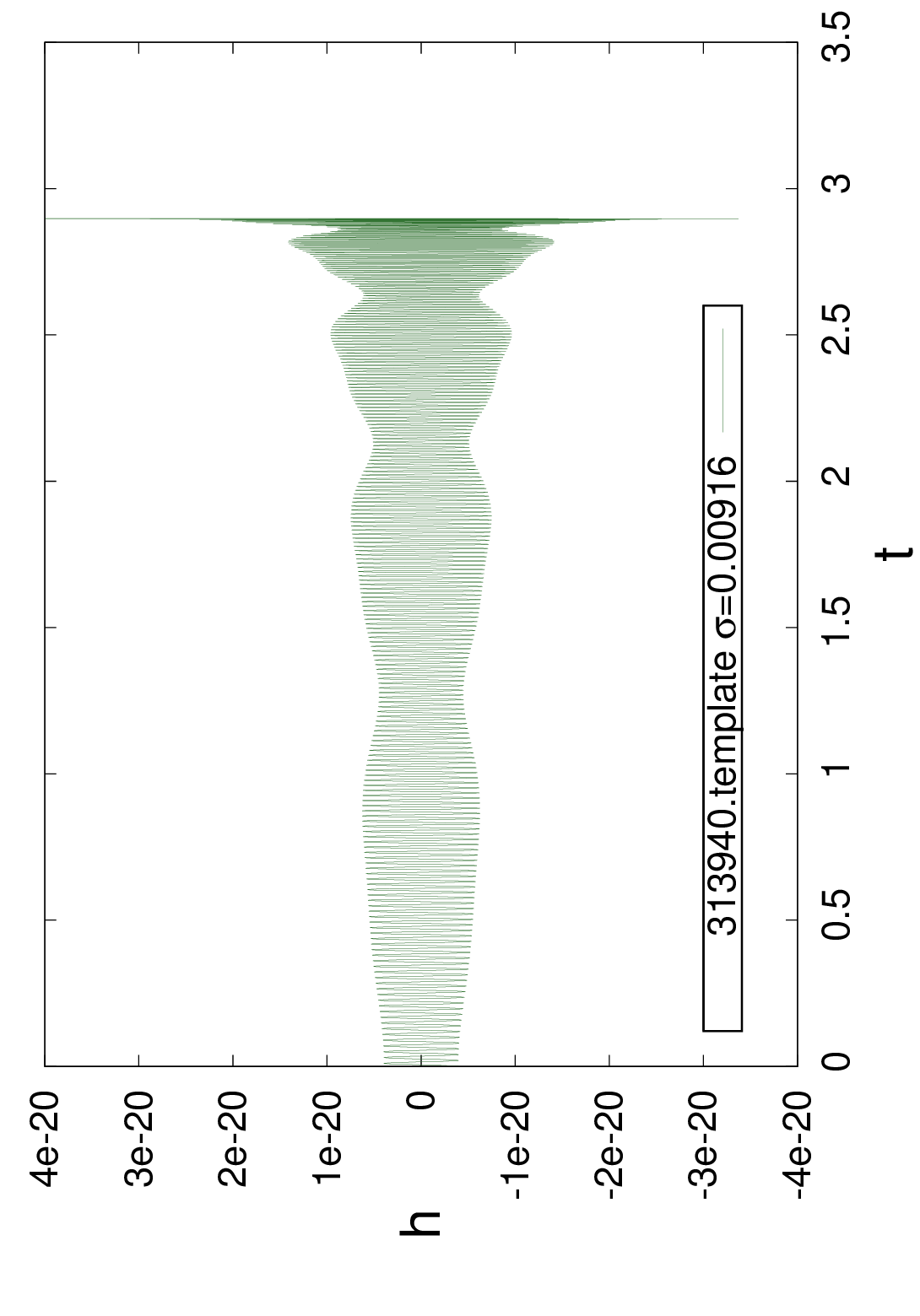} %
\includegraphics[height=8cm, angle=270]{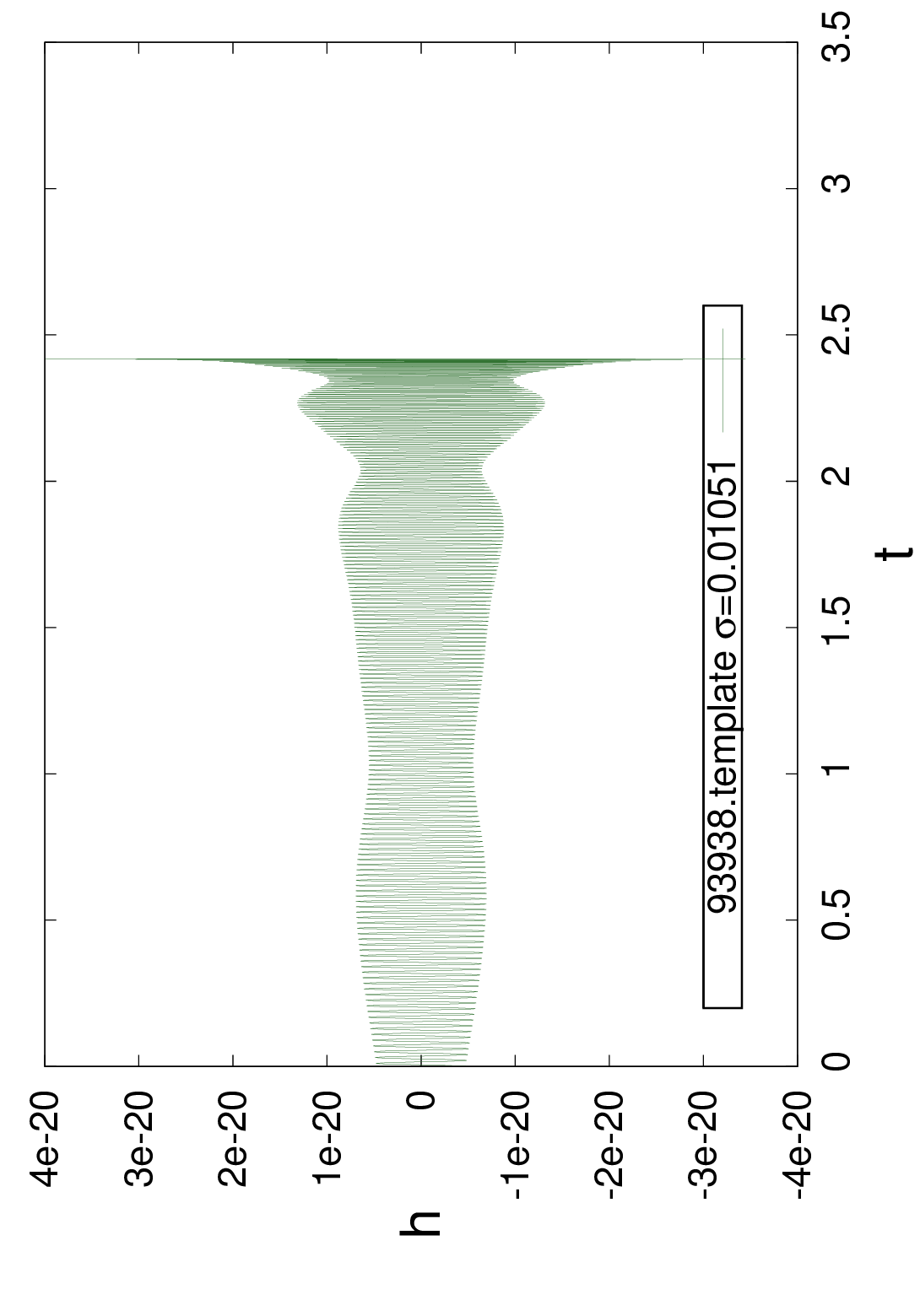} %
\includegraphics[height=8cm, angle=270]{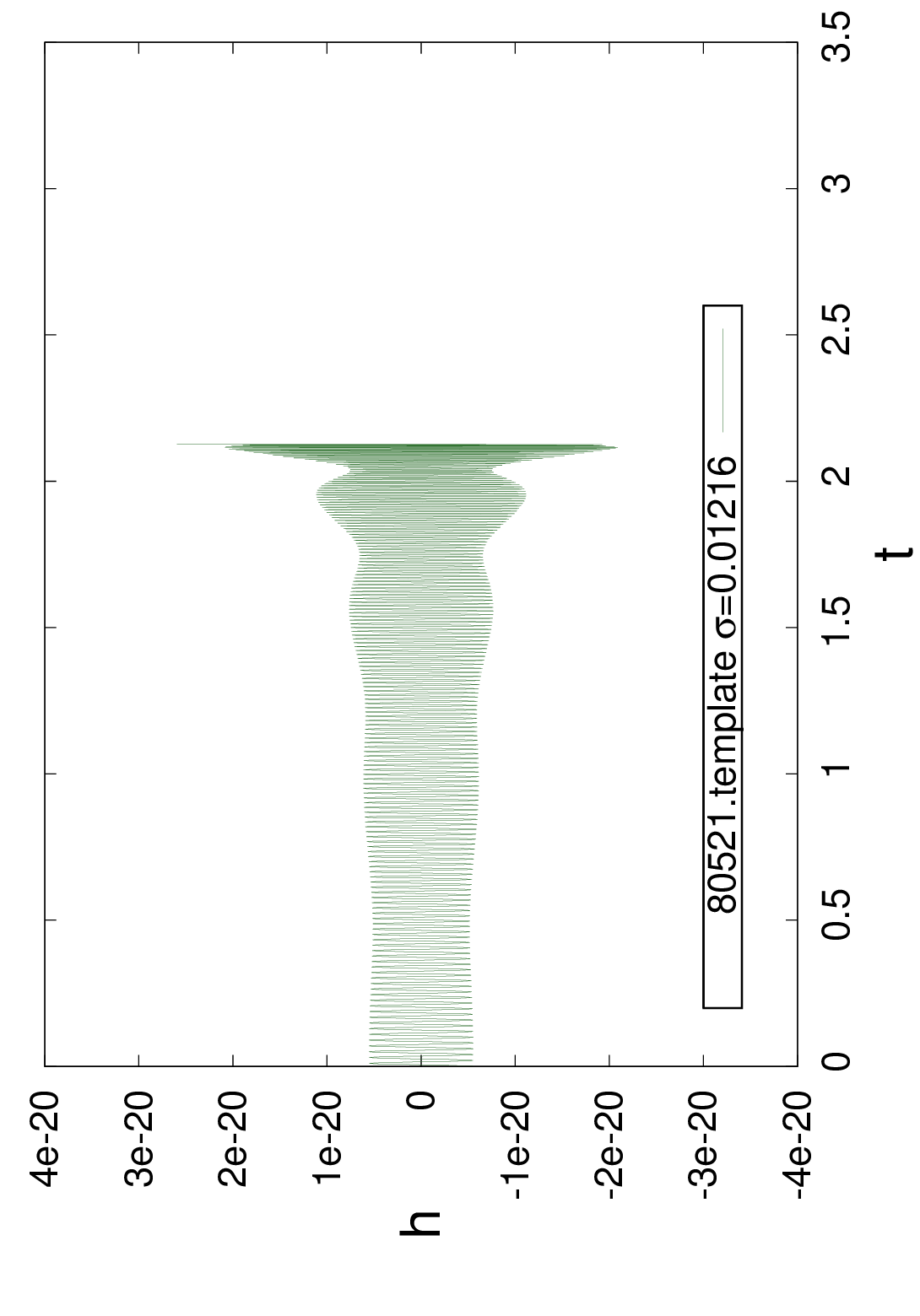}
\caption{ The injected signal and the five best matches. Their serial
numbers from among the $10^{6}$ randomly generated templates is also
indicated.}
\label{fig4}
\end{figure}

The overlap between the noisy injection and the injected signal itself was
found to be $O\left[ h_{n,i},h_{injection}\right] =0.83426$. Therefore we
define an auxiliary quantity%
\begin{equation}
\sigma =\left\vert 1-\frac{O\left[ h_{n,i},h_{template}\right] }{O\left[
h_{n,i},h_{injection}\right] }\right\vert ~
\end{equation}%
for characterizing the closeness of the overlap of a template with the noisy
injection to $O\left[ h_{n,i},h_{injection}\right] $. The parameters of a
number of $>20$ templates with the lowest values of $\sigma $ are
represented on Fig \ref{fig3}. While it is comforting to realize that
(although slightly overestimated) the masses are reasonably well recovered,
the spins are not. There is a considerable scatter in both their magnitudes
and orientations. 
\begin{figure}[th]
\includegraphics[height=3.7cm, angle=0]{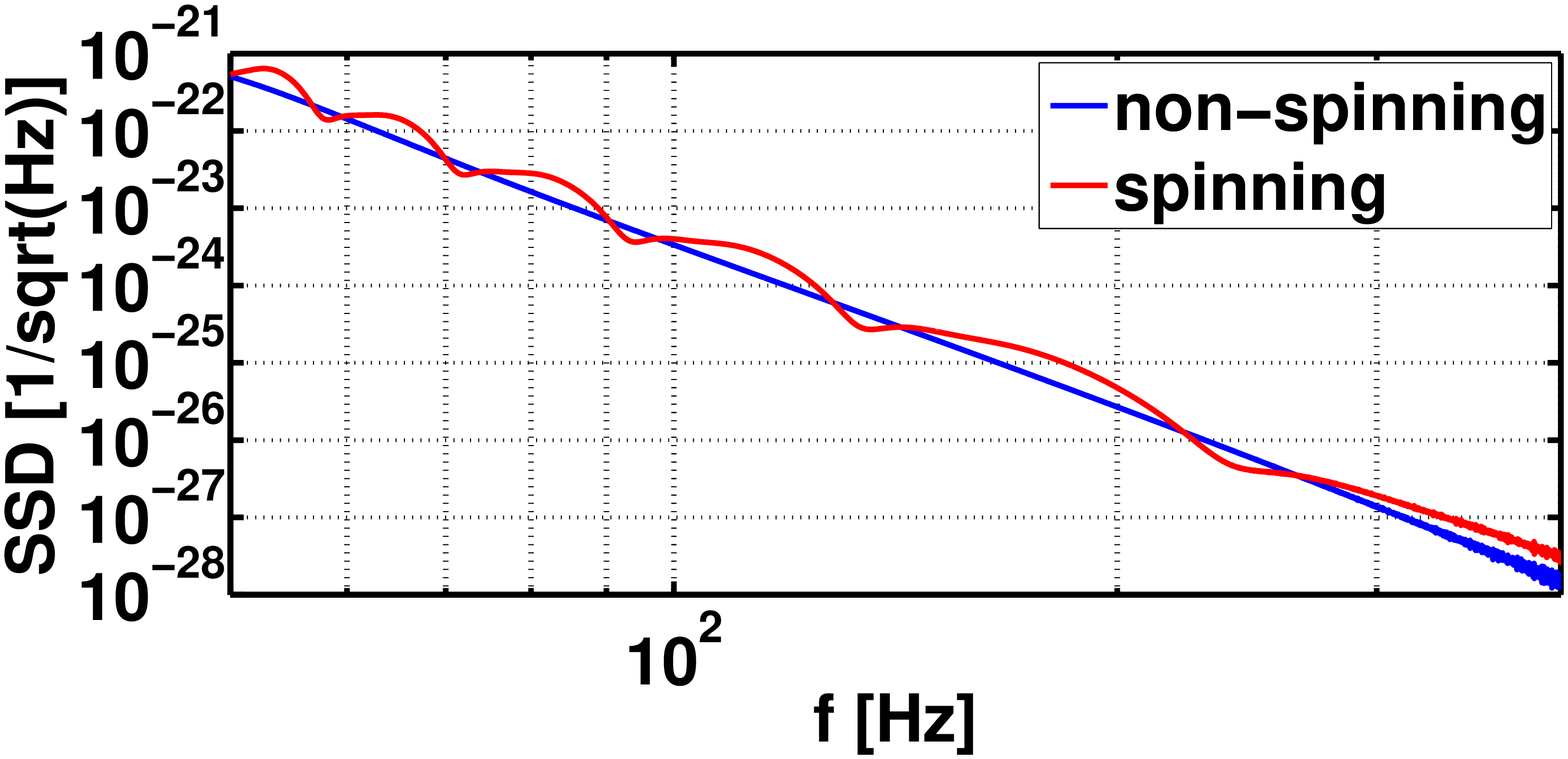} %
\includegraphics[height=3.7cm, angle=0]{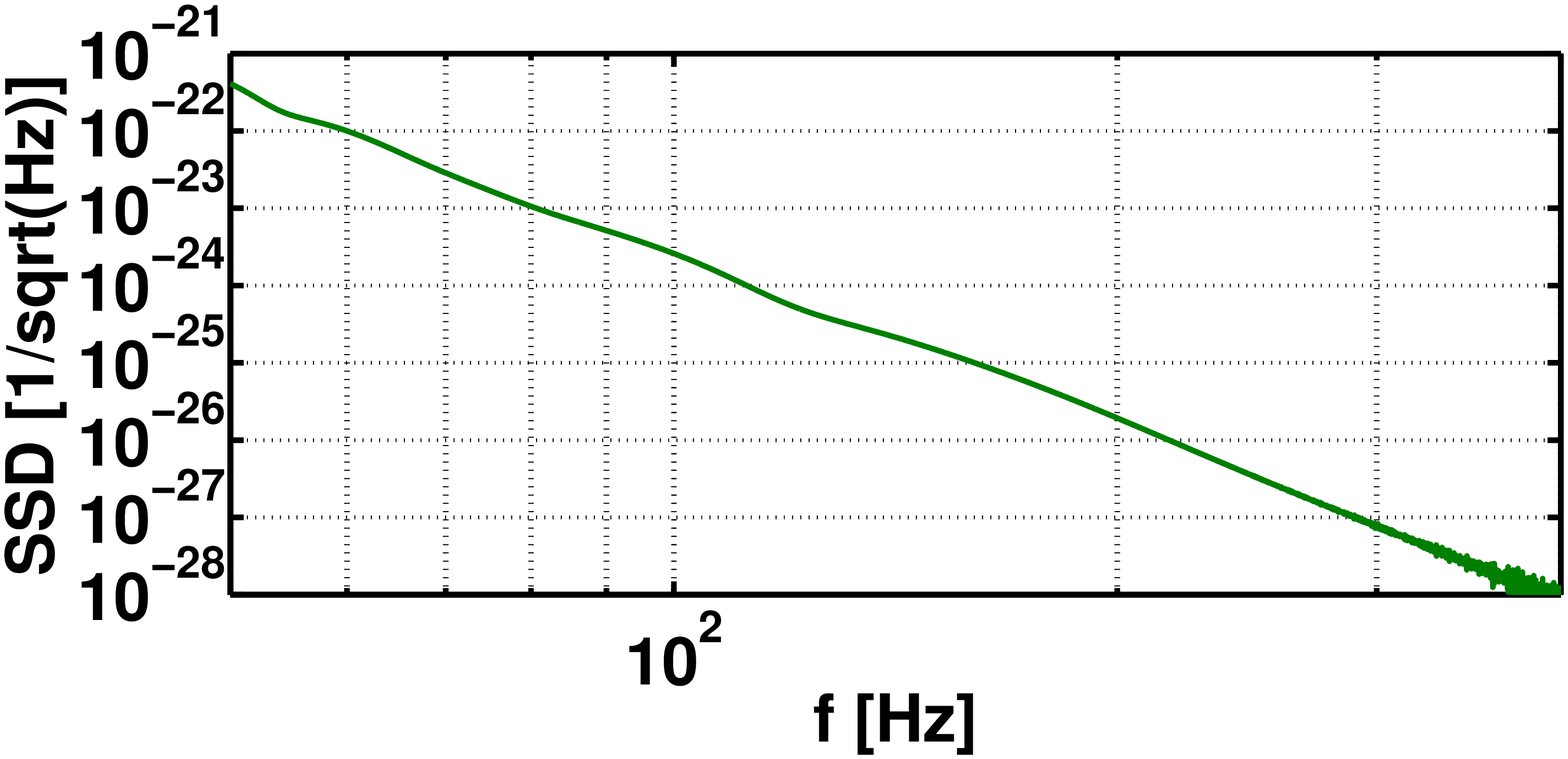} %
\includegraphics[height=3.7cm, angle=0]{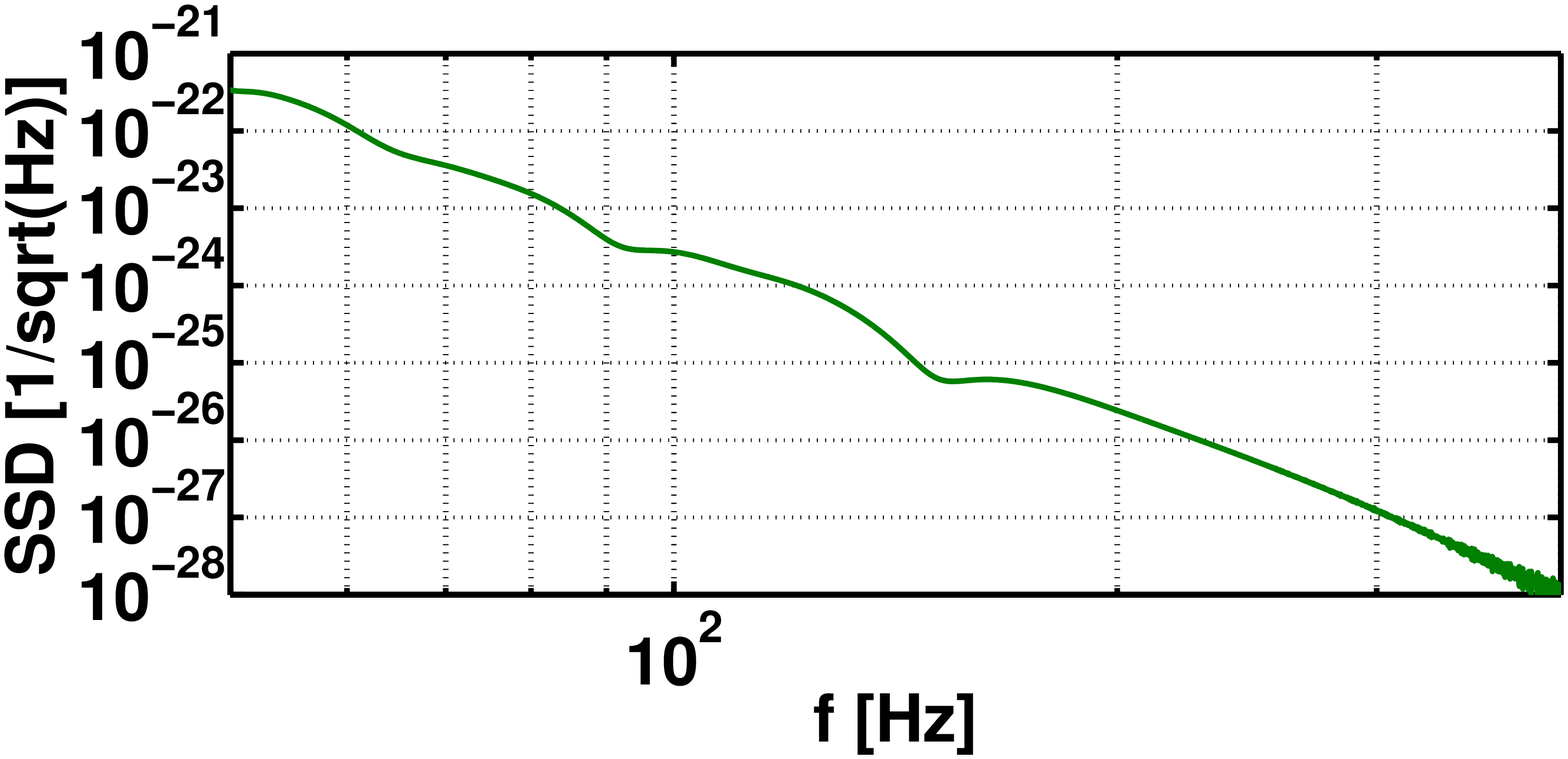} %
\includegraphics[height=3.7cm, angle=0]{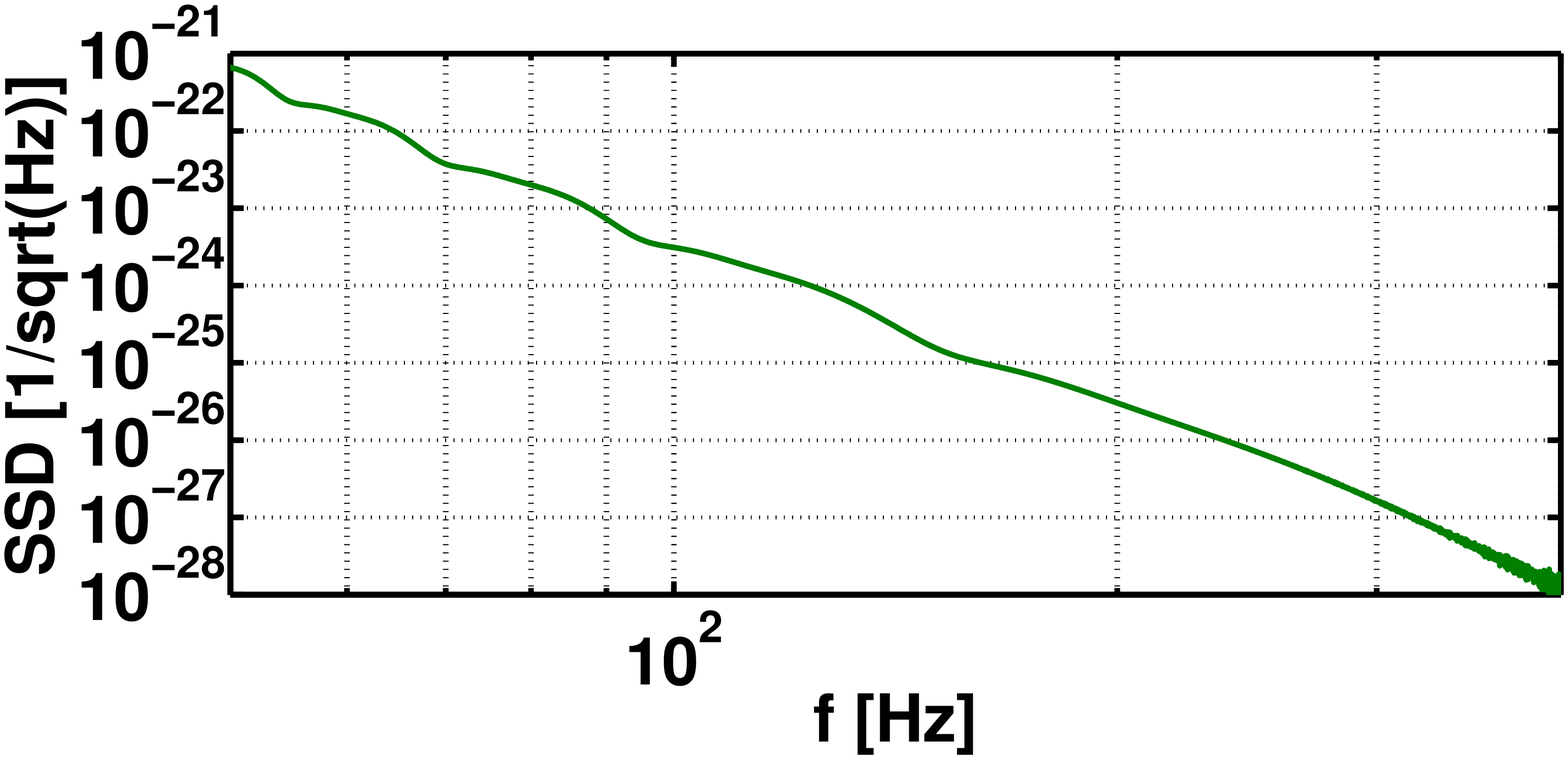} %
\includegraphics[height=3.7cm, angle=0]{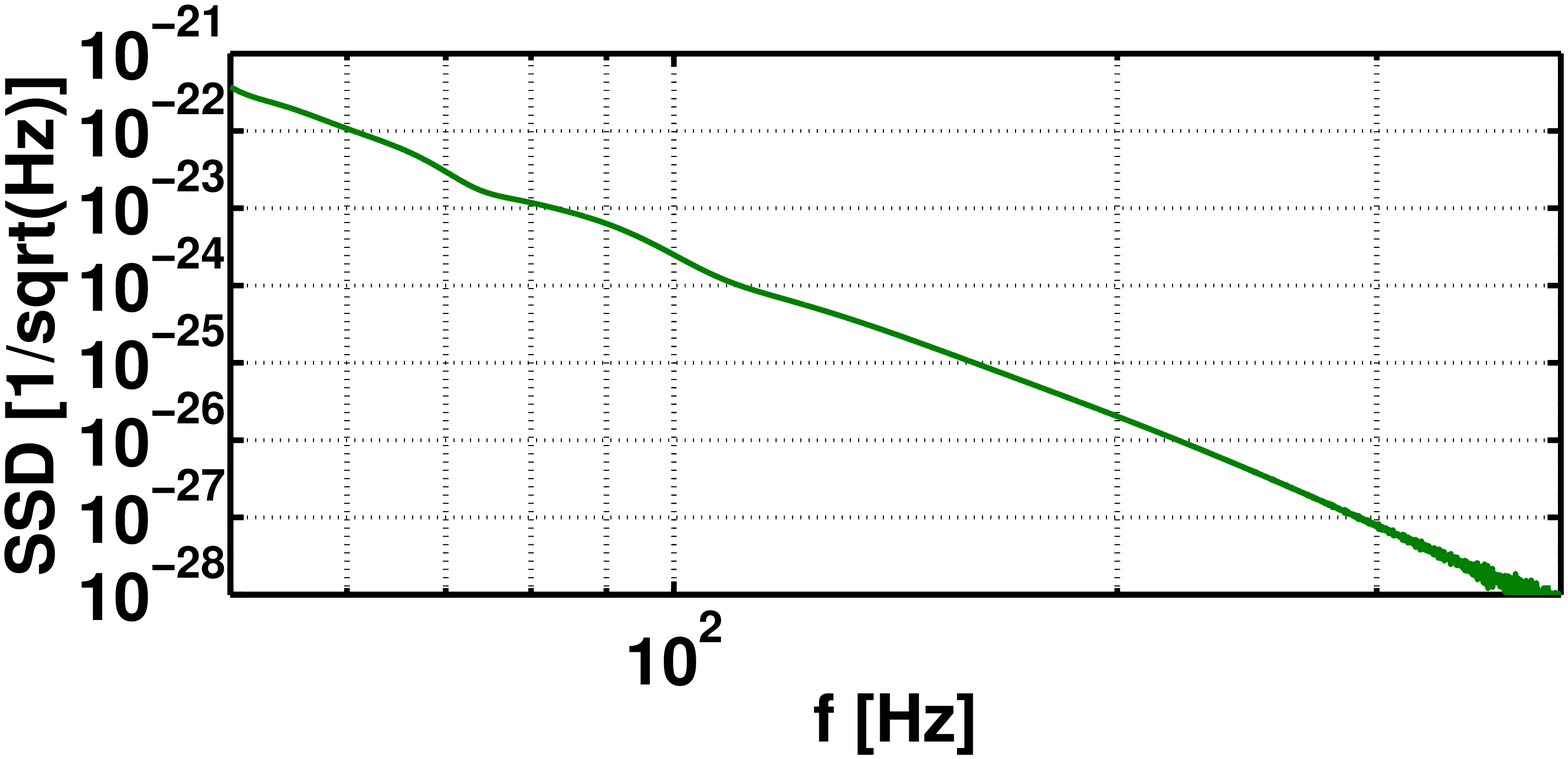} %
\includegraphics[height=3.7cm, angle=0]{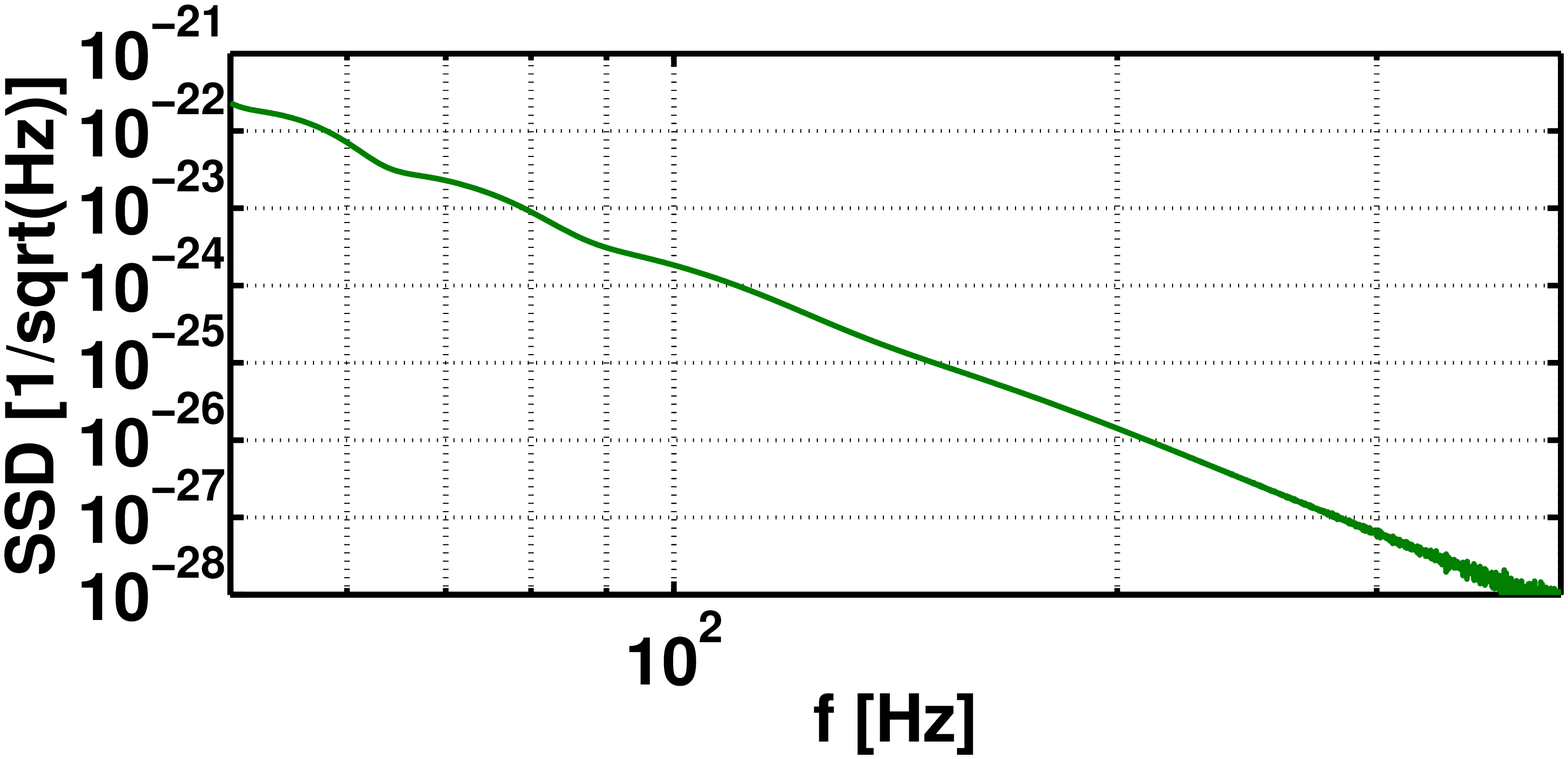}
\caption{ The strain spectral density of the injected signal (red) and its
spin-less counterpart (blue) and the five best matches (green).}
\label{fig5}
\end{figure}

We picked up the 5 best fits, showed their parameters in Table \ref{table1}
and represented the respective waveforms on Fig \ref{fig4}; also their
strain spectral density on Fig \ref{fig5}. For comparison, the injected
signal is also shown.

We conclude that even with all the simplifying assumptions adopted, (a)
either the number of $10^{6}$ templates employed was still insufficient or
(b) the $6$ additional parameters introduced by the spins cause severe
degeneracies, which cannot be resolved by this method. In this context we
note that the use of a Markov-chain Monte-Carlo technique \cite{Vivien}, 
\cite{Vivien2} may turn useful in the recovery of spinning signals. We also
plan to investigate how the use of Advanced LIGO-like data would change
these conclusions.

\textit{Acknowledgements}: We acknowledge interactions with Gyula Szab\'{o}
and Istv\'{a}n Szapudi in the early stages of this work and discussions with
Vivien Raymond on the suitable frequency range. This work was supported by
the Pol\'{a}nyi and Sun Programs of the Hungarian National Office for
Research and Technology (NKTH). ZK and L\'{A}G were further supported by the
Hungarian Scientific Research Fund (OTKA) grant no. 69036. L\'{A}G is
grateful to the organizers of the Amaldi8 meeting for support.

\end{document}